\documentclass[acmsmall,screen,nonacm]{acmart}

\usepackage{algorithm}
\usepackage{algorithmic}
\usepackage{booktabs}
\usepackage{tabularx}
\usepackage{amsmath}
\usepackage{xspace}
\usepackage{xcolor}
\usepackage{multirow}
\usepackage{makecell}
\usepackage{array}
\usepackage{colortbl}
\usepackage{tikz}
\usepackage{scalerel}
\usepackage{fancyvrb}
\usepackage{fvextra}
\usepackage[most]{tcolorbox}

\usetikzlibrary{arrows.meta, positioning, shapes.geometric, backgrounds, fit}

\newcommand{\toolname}{Chronos\xspace}

\newcommand{\openai}{\scalerel*{\includegraphics{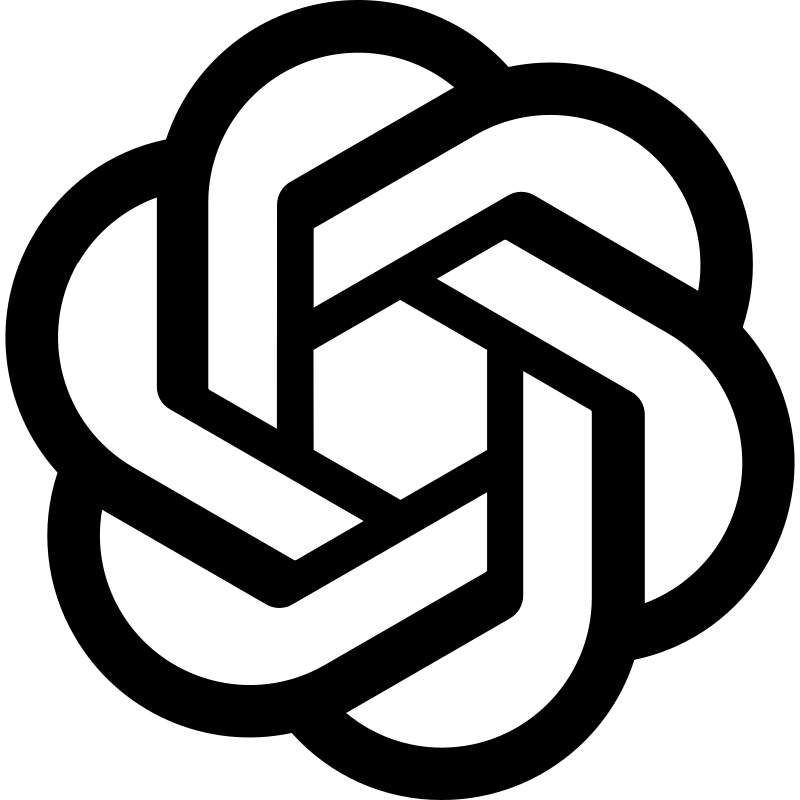}}{\textrm{C}}\xspace}
\newcommand{\anthropic}{\scalebox{1}{\scalerel*{\includegraphics{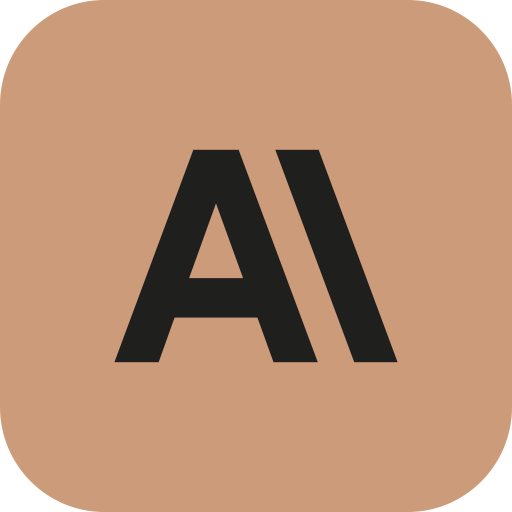}}{\textrm{C}}}\xspace}
\newcommand{\gemini}{\scalebox{1}{\scalerel*{\includegraphics{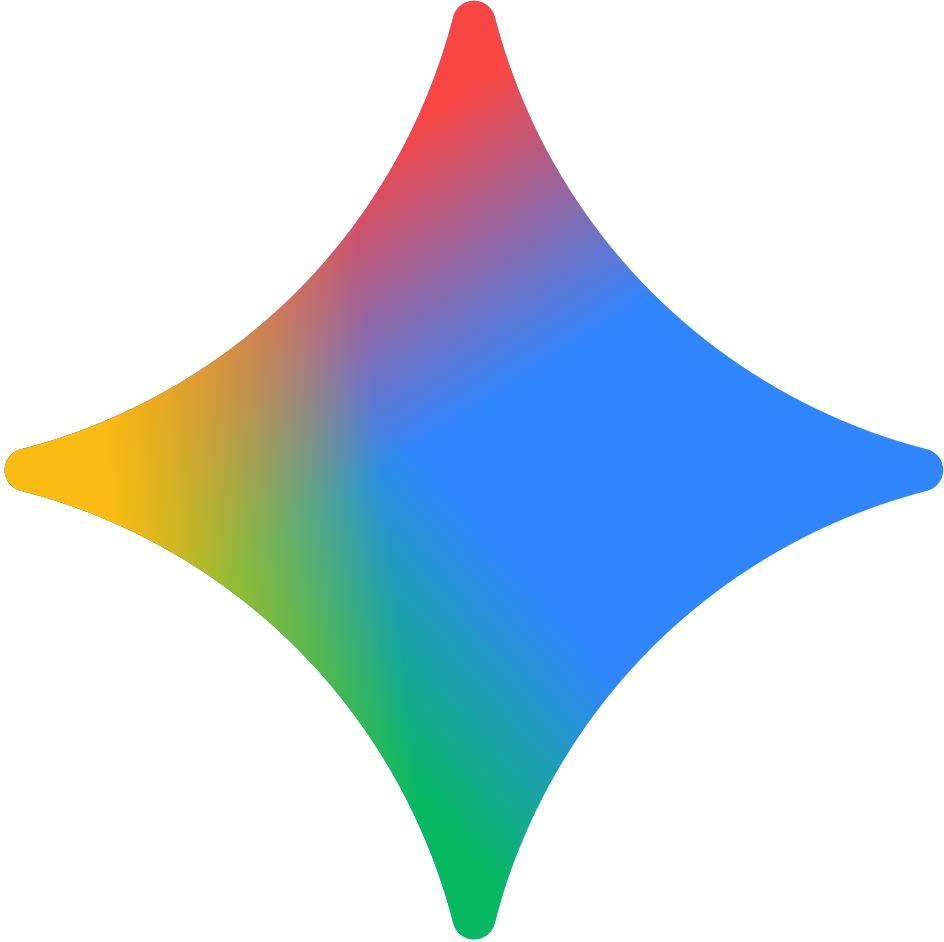}}{\textrm{C}}}\xspace}
\newcommand{\kimi}{\scalerel*{\includegraphics{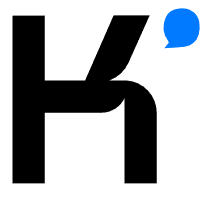}}{\textrm{C}}\xspace}
\newcommand{\qwen}{\scalerel*{\includegraphics{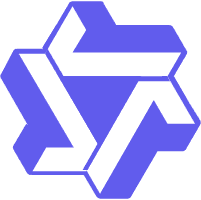}}{\textrm{C}}\xspace}
\newcommand{\minimax}{\scalerel*{\includegraphics{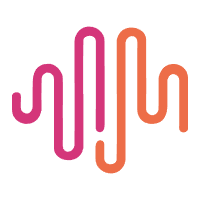}}{\textrm{C}}\xspace}

\definecolor{tier1color}{RGB}{220, 53, 69}
\definecolor{tier2color}{RGB}{255, 193, 7}
\definecolor{tier3color}{RGB}{40, 167, 69}
\definecolor{lightblue}{RGB}{173, 216, 230}
\definecolor{lightgray}{RGB}{245, 245, 245}

\newtcolorbox{promptbox}[1]{
  enhanced,
  breakable,
  lines before break=3,
  boxrule=1.5pt,
  fontupper=\footnotesize,
  fonttitle=\bfseries\color{white},
  arc=5pt,
  rounded corners,
  colframe=blue!50!green,
  colback=blue!5!white,
  coltitle=white,
  title=#1,
  left=4pt
}

\setcopyright{none}
\acmDOI{}
\acmISBN{}

\title[Chronos: Reasoning over Software Evolution]{\toolname Enables Code Agents to Reason over Software Evolution}
\author{Xin Yin}
\authornote{Equal contribution.}
\affiliation{\institution{Zhejiang University}\country{China}}

\author{Yiang Zhang}
\authornotemark[1]
\affiliation{\institution{Zhejiang University}\country{China}}

\author{Zhiyuan Peng}
\affiliation{\institution{Shanghai Jiao Tong University}\country{China}}

\author{Chao Ni}
\affiliation{\institution{Zhejiang University}\country{China}}

\author{Zhe Cui}
\affiliation{\institution{Hithink Research}\country{China}}

\author{Xiaohua Xin}
\affiliation{\institution{National Industrial Information Security Development Research Center}\country{China}}

\authorsaddresses{}
\renewcommand{\shortauthors}{Yin et al.}

\begin{document}

\begin{abstract}
Historical pull requests record the design decisions, compatibility constraints, and implementation patterns behind a codebase's current state. Experience relevant to a new task can span related changes whose descriptions emphasize different concerns. We introduce {\toolname}, a test-time framework that makes this connected history available to large language model (LLM)-based code agents. \toolname distills merged pull requests into structured experience cards and connects them through a typed graph of code-level, developer-intent, and organizational relations. Semantic search identifies entry cards, and weighted multi-hop expansion retrieves connected changes for selective reading. The same memory guides candidate generation and patch selection: a patch-focused change agent and a validation-strategy agent each develop a patch, and an evolution steward consults history to select between them. On SWE-Bench Verified, the full workflow improves SWE-Agent across all six evaluated LLM backbones, raising the mean resolution rate from 69.2\% to 72.9\% and reaching 79.8\% with MiniMax M2.5. With the same backbone, it raises resolution rates from 48.3\% to 51.7\% on SWE-Bench Pro and from 41.0\% to 43.5\% on FEA-Bench Lite. Both experience-guided single-agent variants also outperform the base agent. In a human evaluation on 100 tasks with ten cards retrieved per task, graph-grounded retrieval increases the mean number of useful cards from 1.24 to 2.87 over flat semantic retrieval. These results demonstrate the value of PR relations for retrieving useful repository experience and of the evaluated workflows for applying that experience during patch generation and selection.
\end{abstract}


\maketitle

\begin{figure}[t]
  \centering
  \includegraphics[width=\textwidth]{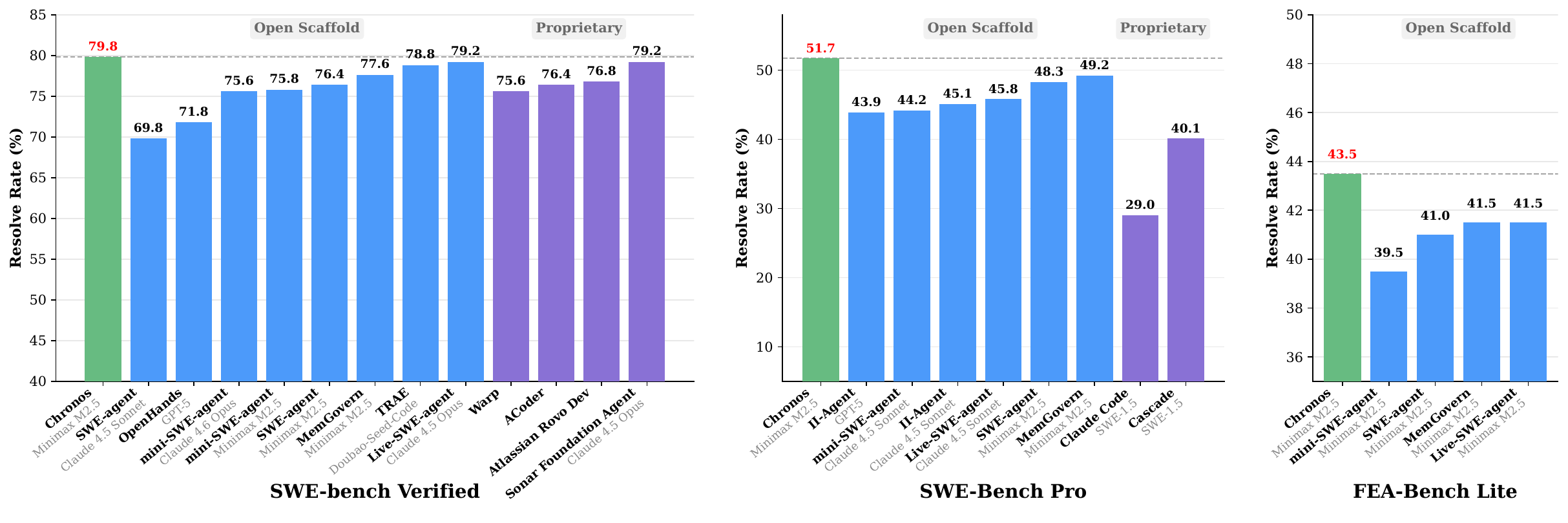}
  \caption{Resolution rates of representative agents across three benchmarks. Labels identify the backbone and scaffold of each configuration; Tables~\ref{tab:main_results} and~\ref{tab:transfer_results} give our backbone-specific comparisons. Each system submits one final patch per task. Colors distinguish \toolname, open-scaffold systems, and proprietary systems.}
  \Description{Comparison of Chronos with open-scaffold and proprietary
  code agents on three software-engineering benchmarks. Colors identify the
  three system groups, and the bars display resolution rates with one final
  patch submitted per task.}
  \label{fig:comparison}
\end{figure}

\section{Introduction}

Code agents based on large language models perform repository-level software engineering tasks by inspecting source code, invoking tools, and producing patches.
SWE-Bench established a common evaluation setting for real-world issue resolution~\cite{jimenez2024swebench}, and frameworks such as SWE-Agent~\cite{yang2024swe} provide tool interfaces for repository exploration and code modification.
Software maintenance also encompasses feature implementation~\cite{li2025fea,zhoufeaturebench}, compatibility maintenance through code migration~\cite{liu2025migrationbench,cheng2025codemenv}, and refactoring~\cite{gautamrefactorbench,thillen2026codetaste}.
Across these tasks, a successful change must satisfy the behavioral requirements and constraints accumulated as a codebase evolves.

The current source tree exposes how a system behaves; its development history records the decisions that shaped that behavior.
Pull requests (PRs) connect code changes with the problems that motivated them, review discussions, and implementation choices.
This context can explain why a compatibility branch exists, which assumption an earlier fix corrected, or how a component was extended after its introduction.
Experienced developers draw on this knowledge when planning new changes.
For a code agent, retrieving the relevant history provides access to prior reasoning that would otherwise have to be reconstructed from the current code.

Recent work makes such experience accessible through memory and retrieval.
MemGovern~\cite{wang2026memgovern} organizes human repair experience into structured cards that agents search and browse, while Repository Memory~\cite{wang2025repomem} uses a target repository's commits and linked issues to support localization.
These approaches make prior solutions available through task-conditioned search and selective inspection.
The organization of that context poses a further challenge: experience relevant to one task can span changes whose descriptions emphasize different concerns.
An initial repair, a subsequent extension, and a related backport may jointly explain how to adapt a component, although only one strongly matches the current query.
Ranking records solely by their similarity to the query leaves these relationships outside the retrieval decision.

Our central insight is that \emph{relations between historical changes provide a second retrieval signal for repository experience}.
PRs expose this signal through repeated changes to the same files, development on newly introduced components, explicit references, and shared workflow context.
These observable relations make connected changes discoverable even when their descriptions emphasize different symptoms or development objectives.
The technical challenge is to follow these relations from semantic entry points, recover additional historical evidence, and prioritize it within a compact shortlist that an agent can inspect.

We introduce \toolname, a test-time framework that turns a repository's PR history into searchable, graph-structured software-evolution experience.
Its memory layer has two complementary parts.
Four-field experience cards capture the \emph{Summary}, \emph{Signals}, \emph{Motivation}, and \emph{Solution} of each historical change.
A typed evolution graph connects those cards through code-level continuity, developer intent, and organizational context.
At task time, semantic search locates entry-point cards, weighted multi-hop expansion finds connected PRs, and the agent selectively reads the resulting experience.
The graph determines which connected changes become candidates for inspection, while the cards explain the rationale and implementation logic needed to apply their experience.

The same memory interface supports both code generation and candidate selection.
Our full resolution workflow uses a \emph{patch-focused change agent} to investigate design intent and a \emph{validation-strategy agent} to investigate behavioral mismatches and regression risks.
Each produces a candidate patch.
An \emph{evolution steward} independently consults the memory and current source code to select the final candidate.
Repository experience thus informs two decisions: how to implement a change and which proposed implementation best addresses the task.
The retrieval service remains separate from these roles and integrates with existing agent tool interfaces without model fine-tuning.

We evaluate \toolname on SWE-Bench Verified~\cite{chowdhury2024verified}, SWE-Bench Pro~\cite{deng2025swe}, and FEA-Bench Lite~\cite{li2025fea}.
On SWE-Bench Verified, the full workflow improves SWE-Agent across all six evaluated LLM backbones, raising the mean resolution rate from 69.2\% to 72.9\% and reaching 79.8\% with MiniMax M2.5.
With the same backbone, \toolname raises resolution rates from 48.3\% to 51.7\% on SWE-Bench Pro and from 41.0\% to 43.5\% on FEA-Bench Lite.
The two single-agent variants reach 78.4\% and 77.8\%, compared with 76.4\% for the base agent, showing that the experience-guided strategies also improve resolution when each produces a single candidate.
In a human evaluation on 100 sampled tasks, graph-grounded retrieval raises the average number of useful cards among ten returned cards from 1.24 to 2.87 and the fraction of tasks receiving useful experience from 68\% to 89\%, compared with flat semantic retrieval.
The benchmark results establish gains for the evaluated resolution workflows, while the retrieval study shows that PR relations help recover useful historical context within a fixed retrieval depth.
Figure~\ref{fig:comparison} places the system results alongside representative agent configurations.

Our contributions are:
\begin{itemize}
\item \textbf{Relational organization of software-evolution experience.} We connect PR-level experience cards through eleven relation types spanning code, developer intent, and organizational context. Semantic entry search and weighted graph expansion use these relations to discover and rank connected historical experience.
\item \textbf{Experience use in generation and selection.} We expose the memory through search and read tools and integrate it into a resolution workflow with two candidate-producing strategies and an independent history-informed selector. The architecture separates reusable repository memory from the agents that consume it.
\item \textbf{Evaluation of task performance and experience retrieval.} We evaluate task resolution on three benchmarks, performance across six backbones on SWE-Bench Verified, and integration with two agent scaffolds. Single-agent and retrieval ablations examine component contributions, while human evaluation measures the useful experience recovered through graph expansion and the effect of removing individual relation types.
\end{itemize}

\begin{figure}[htbp!]
  \centering
  \includegraphics[width=\textwidth]{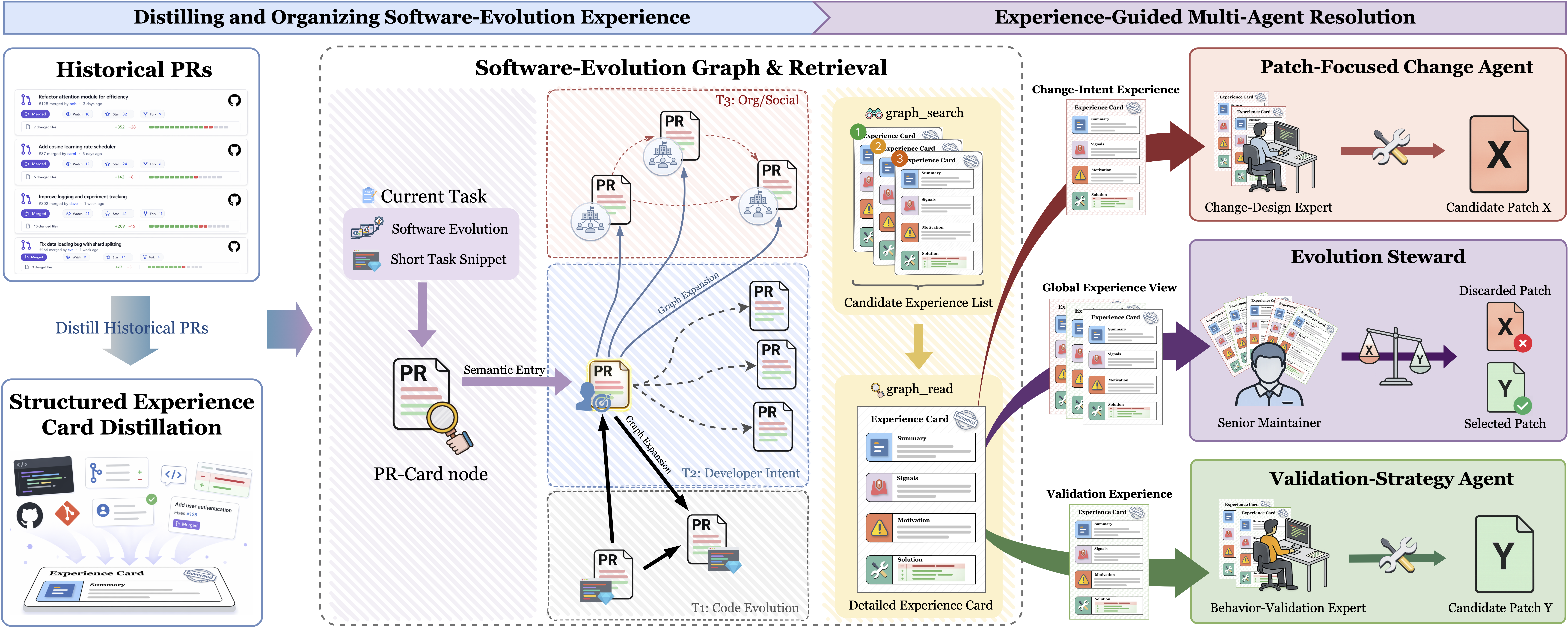}
  \caption{Overview of \toolname. PR-level experience cards are connected by a typed evolution graph. Search and read tools expose this memory to candidate-generating agents and an evolution steward that selects the final patch.}
  \Description{Two-stage Chronos framework. Historical pull requests are distilled into experience cards and connected in a typed evolution graph. Task-time agents retrieve historical experience, generate candidate changes, and select a final patch through an evolution steward.}
  \label{fig:architecture}
\end{figure}

\section{Approach}
\toolname turns repository history into experience that agents can retrieve while investigating a new coding task. Its two stages separate memory construction from task-time use (Figure~\ref{fig:architecture}). The memory layer distills historical PRs into cards describing the rationale and implementation logic of individual changes, then connects those cards through code-level, developer-intent, and organizational relations. At task time, semantic search locates relevant cards and graph expansion retrieves connected changes beyond the semantic shortlist. Agents read selected cards to guide candidate generation and patch selection.

The full workflow combines a patch-focused change agent and a validation-strategy agent with an evolution steward that selects one of their candidate patches. Single-agent variants use the same memory interface with one candidate-producing strategy.

\subsection{Distilling and Organizing Software-Evolution Experience}
\label{sec:construction}

Memory construction produces two linked representations of the same historical PR records. Cards capture the reasoning behind individual changes, and graph edges record the development relations through which those cards can be discovered.

\subsubsection{Software-Evolution Experience Distillation}
\label{sec:cards}

We use merged PRs as historical evolution events and convert each event into a compact, transferable experience card.
A raw PR record mixes code diffs and review discussion with process updates and repository-specific references. Its rationale and implementation logic are dispersed across these sources.
Distillation consolidates that reasoning into a consistent representation that an agent can inspect during its investigation of the current task.

We distill each PR with Qwen3.5-Plus~\cite{qwen35blog} into a fixed four-field card. The fields divide the information needed to find a relevant change from the reasoning needed to adapt it:
\textbf{Summary} states the evolution task and its high-level resolution in one or two sentences and serves as the embedding anchor;
\textbf{Signals} supplies 12--20 concrete retrieval cues covering symptoms or capability gaps, affected component types, APIs, and semantic variants;
\textbf{Motivation} explains why the prior software state was inadequate, tracing triggering inputs or capability gaps to the existing behavior; and
\textbf{Solution} presents the actionable change logic as a development decision, one to three short annotated code patterns, and three to five implementation pitfalls. Search uses the compact Summary and exposes Signals for relevance assessment; reading the full card supplies Motivation and Solution for implementation reasoning.

The distillation prompt takes the PR title, body, discussion, and patch as input and adapts its analysis to bug fixes, feature additions, refactorings, and performance changes. It removes non-technical content and incidental identifiers, including repository and user names, URLs, commit hashes, and version numbers. File paths are allowed only inside Solution code snippets. The original PR identifier links the resulting card to its changed-file metadata and graph node, preserving the source context needed for retrieval. Full prompt templates appear in the supplementary material.

The card and its metadata support different uses of the historical record. The card explains why a change was needed and how it was implemented, so a task query can match its technical concepts without reproducing the source PR's vocabulary. The metadata locates that change within repository development and connects it to earlier and later work. Source identity remains accessible through the PR identifier and linked metadata, even when incidental identifiers have been removed from the embedded text. A card can therefore be discovered through its Summary or through a relation to another PR. After either route, Motivation and Solution help the agent judge whether the underlying assumption and development decision apply to the current task, and how the experience should be adapted to the current code.

\subsubsection{Software-Evolution Graph Construction}
\label{sec:graph}

The second step connects cards using the development relations recorded around their source PRs.
Later changes revisit files, extend newly introduced components, cite previous PRs, and share review or release context.
We construct a directed multigraph $\mathcal{G} = (\mathcal{V}, \mathcal{E})$ over merged PRs, where each node $v \in \mathcal{V}$ represents one historical evolution event and each typed edge records an observable relation between two events.
These links retain development context that abstracted card text omits. Once semantic search identifies a relevant change, retrieval follows links grounded in file continuity, explicit references, or project workflow to find further experience. Table~\ref{tab:edge_types} summarizes the eleven edge types, their detection criteria, observed node coverage, and traversal weights.

We organize these edges into three tiers by evidential strength.
\textbf{Tier 1 (Code Evolution, weight 1.0 to 1.5)} captures directly observable code-level facts: repeated modification of highly overlapping files or continued development on newly introduced files.
\textbf{Tier 2 (Developer Intent, weight 0.7 to 1.05)} encodes explicit intent and short-horizon workflow continuity, including cross-PR references, backports, and same-author follow-ups within nearby directories.
\textbf{Tier 3 (Organizational/Social, weight 0.3 to 0.45)} covers indirect project-process metadata such as shared reviewers, milestones, labels, and comment mentions, which remain informative but are typically noisier.
The traversal weights give larger bonuses to code and workflow continuity and smaller bonuses to broader organizational associations. All detected relations participate in the same bounded expansion; their weights influence candidate scores rather than imposing a fixed traversal order.
We use these hand-specified priors throughout the evaluated settings. Section~\ref{sec:weight_robustness} discusses their role in the memory representation.

\paragraph{Traversal-weight assignment.}
Table~\ref{tab:edge_types} defines the relation criteria, node coverage,
and weight ranges. The weights encode a prior on how much each
relation contributes to a retrieval path. Within an overlap-based or temporal relation,
stronger overlap or shorter separation increases the bonus. The formulas below
specify these assignments, including the 1.0--1.05 overlap that allows a closely
related, temporally nearby follow-up to match or exceed the weakest code-continuity edge.

\begin{table}[t]
  \caption{Software evolution graph edge taxonomy. Coverage reports the fraction of nodes with at least one outgoing edge of each type. Weights are assigned to tiers based on evidential strength.}
  \label{tab:edge_types}
  \centering
  \small
  \setlength{\tabcolsep}{3.5pt}
  \begin{tabularx}{\linewidth}{@{}
    >{\centering\arraybackslash}p{0.055\linewidth}
    >{\raggedright\arraybackslash}p{0.265\linewidth}
    >{\raggedright\arraybackslash}X
    >{\centering\arraybackslash}p{0.095\linewidth}
    >{\centering\arraybackslash}p{0.105\linewidth}
    @{}}
  \toprule
  \textbf{Tier} & \textbf{Edge type (forward / reverse)} & \textbf{Criterion} & \textbf{Cov.} & \textbf{Weight} \\
  \midrule
  \multirow[t]{2}{*}{\textbf{T1}}
    & \texttt{precedes} / \texttt{succeeds} & File-level Jaccard overlap $\geq 0.5$ between PRs. & $\sim$62\% & 1.0--1.5 \\
    & \texttt{extends} / \texttt{extended\_\allowbreak{}by} & A later PR continues development on files newly introduced by an earlier PR. & $\sim$13\% & 1.0 \\
  \midrule
  \multirow[t]{3}{*}{\textbf{T2}}
    & \texttt{follows} / \texttt{followed\_\allowbreak{}by} & Same author, time gap $\leq$ 60 days, and directory Jaccard $\geq 0.5$. & $\sim$73\% & 0.70--1.05 \\
    & \texttt{backport\_\allowbreak{}of} / \texttt{backported\_\allowbreak{}to} & Backport cues detected from title, body, commit messages, or base branch. & $\sim$5\% & 0.70 \\
    & \texttt{references\_\allowbreak{}to} / \texttt{referenced\_\allowbreak{}by} & Explicit \texttt{\#N} citation in PR body or commit messages. & $\sim$46\% & 0.70 \\
  \midrule
  \multirow[t]{6}{*}{\textbf{T3}}
    & \texttt{same\_\allowbreak{}reviewers} & Reviewer overlap (Jaccard $\geq 0.5$). & $\sim$28\% & 0.30--0.45 \\
    & \texttt{same\_\allowbreak{}discussants} & Discussion participant overlap (Jaccard $\geq 0.7$). & $\sim$10\% & 0.30--0.45 \\
    & \texttt{same\_\allowbreak{}milestone} & Shared project milestone. & $\sim$31\% & 0.30 \\
    & \texttt{similar\_\allowbreak{}labels} & Label overlap (Jaccard $\geq 0.5$). & $\sim$30\% & 0.30--0.45 \\
    & \texttt{mentions\_\allowbreak{}in\_\allowbreak{}comments} / \texttt{mentioned\_\allowbreak{}by\_\allowbreak{}in\_\allowbreak{}comments} & Cross-PR mentions through discussion or review comments. & $\sim$17\% & 0.30 \\
    & \texttt{depends\_\allowbreak{}on} / \texttt{dependency\_\allowbreak{}of} & Explicit dependency cues extracted from PR text. & $\sim$0.03\% & 0.30 \\
  \bottomrule
  \end{tabularx}
\end{table}

\paragraph{Tier 1: Code Evolution.}
\textbf{extends}/\textbf{extended\_by} use a fixed weight of $1.0$, because they encode a direct file-creation-to-follow-up-development relation. \textbf{precedes}/\textbf{succeeds} use
\[
w = 0.5 + \texttt{file\_overlap},
\]
where \texttt{file\_overlap} is the file-level Jaccard similarity. The detection threshold of $0.5$ restricts retained edges to similarities in $[0.5,1.0]$, yielding weights in $[1.0,1.5]$.

\paragraph{Tier 2: Developer Intent.}
\textbf{references\_to}/\textbf{referenced\_by} and \textbf{backport\_of}/\textbf{backported\_to} use a fixed weight of $0.7$, because their presence already indicates explicit intent. \textbf{follows}/\textbf{followed\_by} use
\[
w = 0.7 + 0.35 \cdot \frac{t + d}{2},
\]
where
\[
t = 1 - \min(\Delta_{\mathrm{days}}, 60)/60
\]
and directory continuity is normalized as
\[
d = \frac{\min(\max(s_{\mathrm{dir}}, 0.5), 1.0) - 0.5}{0.5},
\]
with $\Delta_{\mathrm{days}}$ the author time gap in days and $s_{\mathrm{dir}}$ the directory Jaccard similarity.
This yields weights in $[0.7, 1.05]$, so nearer follow-up work in more similar directories receives a larger bonus; its upper endpoint overlaps the weakest Tier~1 weight.

\paragraph{Tier 3: Organizational / Social.}
\textbf{same\_\allowbreak{}milestone}, \textbf{mentions\_\allowbreak{}in\_\allowbreak{}comments}, \textbf{mentioned\_\allowbreak{}by\_\allowbreak{}in\_\allowbreak{}comments}, \textbf{depends\_\allowbreak{}on}, and \textbf{dependency\_\allowbreak{}of} use a fixed weight of $0.3$. Three overlap-based relations use small dynamic bonuses above this base value:
\begin{align*}
w_{\mathrm{rev}} &= 0.3 + 0.15 \cdot \frac{\min(\max(s_{\mathrm{rev}}, 0.5), 1.0) - 0.5}{0.5}, \\
w_{\mathrm{dis}} &= 0.3 + 0.15 \cdot \frac{\min(\max(s_{\mathrm{dis}}, 0.7), 1.0) - 0.7}{0.3}, \\
w_{\mathrm{lab}} &= 0.3 + 0.15 \cdot \frac{\min(\max(s_{\mathrm{lab}}, 0.5), 1.0) - 0.5}{0.5},
\end{align*}
where $s_{\mathrm{rev}}$, $s_{\mathrm{dis}}$, and $s_{\mathrm{lab}}$ denote reviewer, discussant, and label Jaccard similarities, respectively.
These formulas yield weights in $[0.3, 0.45]$, giving organizational metadata a smaller per-edge contribution than code-grounded and explicit-intent relations.

\subsection{Retrieving and Applying Software-Evolution Experience}
\label{sec:retrieval}

\subsubsection{Agentic Search over Graph-Grounded Experience}

At task time, agents search and read software-evolution experience as their understanding of the current task develops.
The graph server combines semantic entry search with relation-based expansion, retrieving connected PRs beyond the semantic shortlist.
Two tool calls expose this memory through a shortlist-then-read workflow.
The agent first calls \texttt{graph\_search} to obtain a small set of historical PR candidates, then calls \texttt{graph\_read} on those whose experience appears relevant.
Algorithm~\ref{alg:retrieval} summarizes the bounded retrieval procedure and the filters applied during expansion.

\paragraph{Search experience.}
The \texttt{graph\_search} tool is the agent's retrieval interface.
It accepts a natural-language task summary, optional filename hints, and bounded retrieval parameters, and returns a lightweight shortlist of semantic and graph-expanded candidates.
The server ranks semantic and graph-expanded candidates within their respective groups. The agent-facing shortlist displays each PR's identifier, summary, signals, and a short list of changed files, providing the information needed to assess relevance before reading a full card.
The query $q$ summarizes the current task, and each card is indexed by the embedding of its \textbf{Summary}. We use DashScope \texttt{text-embedding-v4} to compute $\mathbf{e}_q,\mathbf{e}_v \in \mathbb{R}^{2048}$. The index uses cosine distance to return up to $10K$ nearest neighbors. Let $\mathcal{P}_q$ be the candidates that remain after applying the repository, temporal, and optional filename filters described below. The server retains up to $K$ entries in their cosine-similarity order:
\begin{equation}
  \mathcal{S}_0 = \operatorname{TopK}\bigl\{ v \in \mathcal{P}_q \mid \cos(\mathbf{e}_q, \mathbf{e}_v) \bigr\}
\end{equation}
Each entry carries a nonnegative relevance score $a_v=\max\{0,\cos(\mathbf{e}_q,\mathbf{e}_v)\}$. \toolname expands from these entries along typed PR relations, adding connected experience according to both entry relevance and graph evidence. 
Starting from $\mathcal{S}_0$ as the BFS frontier, we expand up to $H$ hops (default $H=3$) with a bounded breadth-first traversal, and score each newly discovered node $u$ reachable from entry $v_0 \in \mathcal{S}_0$ along path $\pi = (v_0, e_1, v_1, \ldots, e_h, u)$ as:
\begin{equation}
  \mathrm{score}(u \mid v_0, \pi)
  = a_{v_0}
  \cdot \exp\!\left(-\frac{h}{\lambda}\right)
  \cdot \Bigl(1 + \sum_{i=1}^{h} w_{e_i}\Bigr)
\label{eq:score}
\end{equation}
where the three factors are entry relevance, a path-length penalty, and an accumulated relation bonus; $h$ is the number of edges on $\pi$, $\lambda = 3.0$ is the decay constant, and $w_{e_i}$ is the edge weight specified in Table~\ref{tab:edge_types}.
The score balances path length against relational evidence: a longer path can rank higher when its accumulated edge weights outweigh the additional length penalty. The hop limit bounds the historical neighborhood explored from each entry.
When multiple explored paths reach $u$, we retain the maximum score. The server returns the semantic entries first, followed by up to $K$ graph-expanded PRs ranked by this score and excluding the entries. The resulting shortlist $\mathcal{R}$ contains at most $2K$ distinct PRs.

The two result groups allocate the shortlist between direct semantic matches and connected historical changes. Reserved positions keep the semantic entries available to the agent while the additional positions broaden the search through PR relations. An expanded node is scored using its originating entry's similarity and the traversed relations, so its own Summary need not rank highly against the query. Expansion can therefore recover a PR outside the initial $10K$ vector candidates, provided an eligible path reaches it. The graph changes the candidate pool through these connections. Ranking the groups separately also avoids treating cosine similarity and a relation-augmented score as interchangeable measures of relevance. Within the expanded group, the maximum over explored paths assigns one score to each PR, even when several entries or relation sequences reach it. Each PR occupies at most one shortlist position, so multiple routes to the same change leave room for other experiences.

\paragraph{Read experience.}
Given a candidate PR identifier, \texttt{graph\_read} returns the complete four-field card for detailed reasoning. The shared identifier links the Summary in the vector index to the full card and graph metadata on the server. The server response retains retrieval scores and paths, while the printed search output exposes the shortlist fields used for relevance assessment. After inspecting a card's Motivation and Solution, the agent can refine its query as source-code investigation clarifies the task. The investigation thus determines which experiences the agent reads in full and whether further retrieval is needed.

\paragraph{Repository and temporal scope.}
Retrieval is scoped to the current task instance. The server derives the
repository from the instance identifier and uses the task PR's creation
time as the cutoff. The same scope applies to semantic entry selection and
graph expansion. A PR is excluded when its
recorded creation or merge time is at or after the cutoff. Filtering both stages
prevents expansion from reintroducing a later change through an otherwise
eligible entry point. Optional filename constraints select PRs that touch an
exact path or a matching path suffix; they constrain both returned semantic
candidates and traversed neighbors. Queries without filenames allow broader
search within the task repository. Filtering the finite semantic candidate
pool can yield fewer than $K$ entries even when additional eligible PRs exist
elsewhere in the index.

\begin{algorithm}[t]
\caption{Bounded graph-grounded experience retrieval}
\label{alg:retrieval}
\begin{algorithmic}[1]
\REQUIRE Graph $\mathcal{G}$, query $q$, task scope, optional filenames,
entry/result limit $K$, hop limit $H$, frontier limit $B$
\ENSURE Ordered shortlist $\mathcal{R}$ with at most $2K$ PRs
\STATE Define $\mathrm{eligible}(v)$ using repository, time, and filename filters
\STATE $\mathcal{P}_q\leftarrow$ eligible PRs among the index's $10K$ nearest candidates
\STATE $\mathcal{S}_0 \leftarrow$ up to $K$ entries from $\mathcal{P}_q$ in semantic order
\STATE $a_v \leftarrow \max\{0,\cos(\mathbf{e}_q,\mathbf{e}_v)\}$ for $v \in \mathcal{S}_0$
\STATE $\mathcal{F} \leftarrow \{(v,[v],a_v,0):v\in\mathcal{S}_0\}$;
       $\mathrm{best}\leftarrow\emptyset$
\FOR{$h=1,\ldots,H$}
  \STATE $\mathcal{F}_{\mathrm{next}} \leftarrow \emptyset$
  \FOR{each state $(u,\pi,a,g)$ in $\mathcal{F}$}
    \FOR{each outgoing typed edge $e=(u,v)$}
      \IF{$v\notin\pi$ and $\mathrm{eligible}(v)$}
        \STATE $g'\leftarrow g+w_e$; $s\leftarrow a\exp(-h/\lambda)(1+g')$
        \IF{$v\notin\mathrm{best}$ or $s>\mathrm{best}[v].\mathrm{score}$}
          \STATE $\pi'\leftarrow\pi\mathbin{\|}[v]$
          \STATE $\mathrm{best}[v]\leftarrow(s,\pi',a,g')$
          \STATE Append $(v,\pi',a,g')$ to $\mathcal{F}_{\mathrm{next}}$
        \ENDIF
      \ENDIF
    \ENDFOR
  \ENDFOR
  \STATE $\mathcal{F}\leftarrow$ up to $B$ next states ranked by their node's best score
  \IF{$\mathcal{F}=\emptyset$}
    \STATE \textbf{break}
  \ENDIF
\ENDFOR
\STATE $\mathcal{S}_g\leftarrow$ top-$K$ nodes in $\mathrm{best}$ excluding $\mathcal{S}_0$
\RETURN $\mathcal{S}_0\mathbin{\|}\mathcal{S}_g$
\end{algorithmic}
\end{algorithm}

\paragraph{Traversal state and search budget.}
Each frontier state carries the current PR, the path used to reach it, the
originating semantic score, and the accumulated edge-weight bonus. Expansion
preserves the entry score while updating the path length and graph bonus in
Equation~\ref{eq:score}. A node already on the current path is skipped, so a
cycle cannot accumulate weight by repeatedly revisiting the same PR. The server
records each newly reached candidate and updates its record when a path improves
the stored score. Traversal stops when no frontier remains or the hop limit is
reached. In our implementation, each subsequent frontier is capped at $B=200$
states. This limits how many states continue through densely connected regions,
prioritizing states by their nodes' highest scores among the explored paths.

Relation types remain distinct during traversal. Because $\mathcal{G}$ is a
multigraph, the same ordered pair of PRs can be connected by several edge
types. Each outgoing typed edge supplies an alternative extension of the
current path, and Equation~\ref{eq:score} uses the weight of that edge.
The relation bonus therefore sums weights along a path, rather than summing
all metadata associations between each pair of adjacent PRs. For parallel
edges extending the same frontier state, the larger weight produces the
higher candidate score when the entry relevance is positive. The stored
maximum retains that score, while the PR still occupies one result position.
Different forms of historical evidence thus provide alternative
retrieval routes without accumulating bonuses for several relation types
on the same connection. Multi-hop bonuses require
traversing successive PRs within the eligible historical neighborhood.

The hop and frontier limits bound graph exploration, while the $2K$ result
limit bounds the shortlist exposed to the agent. Expansion preserves the
semantic entries and allocates the additional $K$ positions to connected
history. Repository, temporal, and filename filters can further reduce the
number of returned PRs.

\subsubsection{Multi-Agent Collaboration over Software Evolution}
\label{sec:ensemble}

The full workflow generates two candidates in separate agent runs, then selects between them through an independent review informed by repository history. Both candidate agents receive the current task and repository. They are instructed to inspect relevant source files, reproduce the behavior gap, modify the source code, rerun the reproduction, and consider related edge cases. Their roles differ in how they formulate retrieval queries and judge the applicability of a precedent. Full role prompts appear in the supplementary material.

\textbf{Patch-focused change agent.}
This agent formulates queries around design intent, including the broken invariant, violated assumption, and desired behavior. It reads relevant cards to understand the reasoning behind prior changes and assesses whether that reasoning addresses the same kind of logical flaw. Experience from the same module is a strong relevance signal. Experience from another module can also guide implementation when the underlying invariant or assumption transfers. The agent further considers whether the precedent's implementation complexity is appropriate for the current task.

\textbf{Validation-strategy agent.}
This agent formulates queries around differences between expected and actual behavior, emphasizing concrete symptoms, affected APIs, and regression risks. It consults history to resolve uncertainty about unfamiliar APIs or complex changes, reading cards about similar files or the same type of behavioral issue. To assess applicability, it checks whether the experience concerns the same module and behavioral mismatch, and whether the proposed complexity is appropriate. Experience from another module primarily serves as a conceptual reference. Both candidate prompts encourage agents to stop retrieving once one or two relevant experiences provide a clear direction and to adapt the retrieved reasoning to the current source code.

\textbf{Evolution steward.}
For distinct, non-empty candidate patches $C_x$ and $C_y$, the steward receives the task description and both diffs. Their X/Y order is randomized when review instances are constructed to reduce position bias. The steward can inspect current source code and independently retrieve cards from both change-intent and validation perspectives. It analyzes each patch's modifications, treatment of the root cause, consistency with historical experience, and missed or newly introduced edge cases. The decision criterion is logical correctness, excluding patch length, style, and comment density. Calling \texttt{choose\_patch x} or \texttt{choose\_patch y} records the verdict and ends the session. The steward selects an existing candidate without editing files or generating a third patch.

Historical experience informs a different decision at each stage. During generation, it helps an agent determine what to change and which behavior to preserve. During selection, the candidate diffs make those questions concrete: the steward compares proposed modifications with relevant historical rationale and examines whether that rationale applies to the affected code. Its independent search can use the candidate content to guide retrieval beyond the cards consulted by either generator. The final choice is grounded in the current task and source code, with history supplying development constraints for comparing the two available implementations.

The component ablations in Section~\ref{sec:ablation} retain one candidate-producing perspective and bypass two-candidate selection. The routing procedure below handles identical or empty outputs before invoking the steward.

\paragraph{Candidate routing and final patch.}
Table~\ref{tab:patch_routing} defines how candidate outputs enter the
selection stage. Cases with identical patches or fewer than two non-empty
patches are resolved directly; distinct, non-empty patches enter steward
review. Each task submits one selected patch.

\begin{table}[t]
\caption{Routing of candidate patches before final submission. Equality is
checked after stripping surrounding whitespace.}
\label{tab:patch_routing}
\centering
\small
\begin{tabularx}{\linewidth}{@{}lX@{}}
\toprule
\textbf{Candidate outputs} & \textbf{Selection action} \\
\midrule
Two identical, non-empty patches & Retain the shared patch directly. \\
One non-empty patch & Retain the available patch directly. \\
Two empty patches & Submit an empty patch. \\
Two distinct, non-empty patches & Randomize their X/Y order and invoke the steward. \\
\bottomrule
\end{tabularx}
\end{table}

Each review instance retains the mapping from X/Y to the originating strategy,
so the recorded decision identifies the corresponding original candidate.
If a review session produces no valid X/Y choice, the implementation selects
one of the two existing candidates at random using a fixed seed.
The benchmark then evaluates the selected patch using the tests for the current
task.

\section{Experimental Setup}
\label{sec:setup}

\subsection{Agents and Retrieval Configuration}
We integrate \toolname's structured software-evolution experience and
graph-grounded search into SWE-Agent~\cite{yang2024swe} and compare the resulting
workflow with vanilla SWE-Agent. We also implement a mini-SWE-Agent variant,
denoted \toolname$_{\text{mini}}$, to examine integration with a second scaffold.
The same retrieval-tool interface is in principle compatible with other
scaffolds, including OpenHands~\cite{wangopenhands}.

Unless otherwise noted, each agent run has a limit of 250 steps, a cost
budget of \$1.5, and a context window of 200k tokens. The full workflow includes
two candidate-generation runs and invokes the steward for distinct, non-empty
candidates, as specified in Table~\ref{tab:patch_routing}. The graph server runs
as a persistent service, with $K=5$ semantic entry points and $H=3$ expansion hops.
Each query returns a shortlist of at most ten experience cards. On SWE-Bench
Verified, we evaluate GPT-5 Mini, Claude 4.5 Haiku, Gemini 3 Flash, Kimi K2.5,
Qwen3-Max, and MiniMax M2.5. The two transfer benchmarks use MiniMax M2.5,
the strongest evaluated backbone on SWE-Bench Verified.

Historical retrieval is scoped to each target instance. The server resolves the
repository and temporal cutoff from the instance identifier and excludes PRs whose
recorded creation or merge timestamp is at or after that cutoff.
These filters apply to semantic entry selection and graph expansion
(Algorithm~\ref{alg:retrieval}). Each workflow submits one final patch per task.
Section~\ref{sec:variance} describes independent repetitions of the full
configuration.

\subsection{Benchmarks and Task Coverage}
\label{sec:benchmarks}
We evaluate issue resolution and feature implementation on the three benchmark
subsets in Table~\ref{tab:benchmarks}.

\begin{table}[t]
\caption{Benchmark subsets used in the evaluation, with task and repository counts.}
\label{tab:benchmarks}
\centering
\small
\begin{tabularx}{\linewidth}{@{}lccX@{}}
\toprule
\textbf{Benchmark} & \textbf{Tasks} & \textbf{Repos.} & \textbf{Task emphasis} \\
\midrule
SWE-Bench Verified & 500 & 12 & Human-verified issue resolution \\
SWE-Bench Pro (public) & 731 & 11 & Longer-horizon repository-level changes \\
FEA-Bench Lite & 200 & 48 & New feature implementation \\
\bottomrule
\end{tabularx}
\end{table}

\paragraph{SWE-Bench Verified.}
SWE-Bench Verified~\cite{chowdhury2024verified} is a human-filtered subset of
500 SWE-Bench instances. Human review screens for well-specified issue
descriptions and appropriately scoped unit tests. We use it as the
primary issue-resolution benchmark and evaluate all six backbones on it.

\paragraph{SWE-Bench Pro.}
SWE-Bench Pro~\cite{deng2025swe} targets longer-horizon changes in large
codebases. Its full collection contains 1,865 tasks from 41 repositories; the
public setting used here contains 731 tasks from 11 repositories. Human-augmented
requirements and interface specifications clarify the expected behavior.
This setting tests experience reuse in changes that require more extensive
coordination across files and implementation constraints.

\paragraph{FEA-Bench Lite.}
FEA-Bench~\cite{li2025fea} derives feature-implementation tasks from PRs and
pairs them with unit tests. The full dataset has 1,401 tasks from 83 repositories;
its Lite subset has 200 tasks from 48 repositories. We use the Lite subset to
evaluate the addition of new components together with edits to existing code.
These tasks test the integration of new behavior into the repository's
established structure.

Resolution rate is the fraction of target tasks
solved by the submitted patches under the corresponding benchmark evaluation.
Across benchmarks, we retain the graph-construction rules, card schema, and
default weight formulas without benchmark-specific retuning. The historical
content is scoped to the repository of each target task.

\subsection{Human Evaluation of Retrieved Experience}
\label{sec:human_protocol}

We conduct an internal expert annotation study to assess whether retrieved
cards provide useful guidance for the current task. We randomly sample 100
SWE-Bench Verified instances and retrieve the top-ten cards for each task under
Random, Embedding, and \toolname. Random samples cards uniformly from the
knowledge base, Embedding ranks cards by cosine similarity, and \toolname uses
graph-grounded search. All methods use the same task sample and retrieval depth.

\paragraph{Annotation inputs and rubric.}
Two members of our research team with professional software-engineering
experience independently label each retrieved card using the task description
and all four card fields: Summary, Signals, Motivation, and Solution. A card is
\emph{useful} when it identifies a transferable implementation pattern, failure
mode, compatibility constraint, design rationale, or validation strategy that
can directly help produce or check a patch for the current task. The remaining
cards are \emph{not useful}; topical similarity or a matching identifier alone
does not establish actionable guidance.

Annotators focus on transferable root causes, implementation constraints,
affected components, and implications for validation, without consulting
external information. Usefulness is task-conditioned: a card from a different
component can be useful when its reasoning applies to the current task.

\paragraph{Agreement and reported metrics.}
Disagreements are resolved through discussion. Inter-annotator agreement before
adjudication is Cohen's $\kappa=0.86$.
For task $i$ and retrieved card $j$, let $u_{ij}=1$ for an adjudicated useful
card and $u_{ij}=0$ otherwise. With $N=100$ tasks and $m=10$ cards per task,
we report
\begin{align}
\mathrm{Useful} &= \frac{1}{N}\sum_{i=1}^{N}\sum_{j=1}^{m}u_{ij}, \\
\mathrm{Hit} &= \frac{1}{N}\sum_{i=1}^{N}
  \,\mathbf{1}\!\left\{\textstyle\sum_{j=1}^{m}u_{ij}>0\right\}.
\end{align}
Here, $\mathbf{1}\{\cdot\}$ denotes the indicator function.
Useful measures the mean number of useful cards per task, while
Hit measures the fraction of tasks receiving at least one useful card.
The edge-type ablation uses the same annotation protocol to measure changes in Useful.

\section{Evaluation}

\subsection{Main Results}

\textbf{SWE-Bench Verified.}
Table~\ref{tab:main_results} reports resolution rates across six LLM backbones.
The full \toolname workflow improves SWE-Agent on all six, raising the mean resolution rate from 69.2\% to 72.9\%, a gain of $3.7$ percentage points (pp).
Qwen3-Max gains $+6.2$pp, and the other backbones gain $+2.8$pp to $+3.6$pp.
Gains occur with both the lowest-scoring base configuration, GPT-5 Mini, and the highest-scoring one, MiniMax M2.5, showing that the workflow improves performance across the evaluated range of backbones.
The analyses below examine run-to-run stability, integration with different scaffolds, and the contributions of candidate generation and retrieval.

\begin{table}[htbp!]
  \caption{Resolution rates (\%) on SWE-Bench Verified. $\Delta$ is the gain in percentage points.}
  \label{tab:main_results}
  \small
  \centering
  \begin{tabular}{lccc}
  \toprule
  \textbf{LLM Backbone} & \textbf{SWE-Agent} & \textbf{\toolname} & \textbf{$\Delta$} \\
  \midrule
  \openai GPT-5 Mini        & 57.2\% & \textbf{60.4\%} & $+3.2$ \\
  \anthropic Claude 4.5 Haiku  & 65.6\% & \textbf{68.4\%} & $+2.8$ \\
  \gemini Gemini 3 Flash    & 75.8\% & \textbf{78.6\%} & $+2.8$ \\
  \kimi Kimi K2.5         & 71.2\% & \textbf{74.8\%} & $+3.6$ \\
  \qwen Qwen3-Max         & 69.0\% & \textbf{75.2\%} & $+6.2$ \\
  \minimax MiniMax M2.5      & 76.4\% & \textbf{79.8\%} & $+3.4$ \\
  \midrule
  \multicolumn{1}{l}{\textit{Average}} & 69.2\% & 72.9\% & $+3.7$ \\
  \bottomrule
  \end{tabular}
\end{table}

\textbf{SWE-Bench Pro and FEA-Bench Lite.}
With MiniMax M2.5, \toolname raises the resolution rate from 48.3\% to 51.7\% ($+3.4$pp) on SWE-Bench Pro and from 41.0\% to 43.5\% ($+2.5$pp) on FEA-Bench Lite (Table~\ref{tab:transfer_results}).
These gains extend the result from standard issue resolution to longer-horizon repair and feature implementation.
The graph-construction rules and default weight formulas remain unchanged across benchmarks, with historical content drawn from each target repository (Section~\ref{sec:benchmarks}).
The same memory-construction and retrieval procedures therefore support distinct development tasks without benchmark-specific retuning.

\begin{table}[htbp!]
  \caption{Resolution rates (\%) on the transfer benchmarks using MiniMax M2.5. $\Delta$ is the gain in percentage points.}
  \label{tab:transfer_results}
  \small
  \centering
  \begin{tabular}{lccc}
  \toprule
  \textbf{Benchmark} & \textbf{SWE-Agent} & \textbf{\toolname} & \textbf{$\Delta$} \\
  \midrule
  SWE-Bench Pro & 48.3 & \textbf{51.7} & $+3.4$ \\
  FEA-Bench Lite & 41.0 & \textbf{43.5} & $+2.5$ \\
  \bottomrule
  \end{tabular}
\end{table}

\subsection{Run-to-Run Stability}
\label{sec:variance}

We repeated the full \toolname configuration with the SWE-Agent scaffold and MiniMax M2.5 five times over all 500 SWE-Bench Verified instances.
The graph, retrieval parameters, prompts, step limits, and per-agent cost limits remain fixed across runs; backbone sampling varies between repetitions.
The resolution rates are 79.8\%, 80.0\%, 80.6\%, 79.6\%, and 80.0\%, giving a mean of 80.0\%, a sample standard deviation of 0.37pp, and a standard error of 0.17pp.
The 95\% Student's $t$ interval for this configuration's mean is $[79.5\%,80.5\%]$.
The five runs span 1.0pp, and the lowest result, 79.6\%, remains 3.2pp above the reported SWE-Agent reference of 76.4\%.

Table~\ref{tab:main_results} reports the first \toolname run, 79.8\%.
Each repetition submits its own final patches, without pooling candidates across runs.
The baseline and other configurations use single runs. The interval therefore characterizes the mean performance of the repeated full configuration, rather than the difference between methods.

\subsection{Ablation}
\label{sec:ablation}
Table~\ref{tab:ablation} examines \toolname with MiniMax M2.5 along three dimensions: integration with two scaffolds, operation with one or two candidate-producing strategies, and graph-grounded versus flat semantic retrieval.
The single-agent variants retain graph-grounded experience search and bypass the steward, so each produces one candidate patch.

\textbf{Scaffold integration.}
\toolname raises SWE-Agent from 76.4\% to 79.8\% and mini-SWE-Agent from 75.8\% to 79.2\%, yielding the same $+3.4$pp gain under both scaffolds.
These results show that the same memory interface supports two agent implementations while preserving the graph-construction and retrieval procedures.

\textbf{Single-agent use and candidate selection.}
The change-agent variant reaches 78.4\%, and the validation-agent variant reaches 77.8\%.
Each exceeds the 76.4\% SWE-Agent baseline while generating one candidate without steward selection.
The gains therefore extend to both experience-guided strategies in single-candidate operation.
Combining the two strategies with the steward reaches 79.8\%, an additional $1.4$pp over the change-agent variant and $2.0$pp over the validation-agent variant.
The full workflow achieves the highest resolution rate among the evaluated configurations.

\textbf{Relational experience retrieval.}
The full workflow reaches a resolution rate of 79.8\% with graph-grounded retrieval and 79.0\% with flat semantic search, a gain of $+0.8$pp.
The retrieval study in Section~\ref{sec:graph_analysis} examines how this component changes the experience available to the agents.
The human study measures the usefulness of retrieved cards, while this ablation measures the effect of retrieval choice within the full coding workflow.

\begin{table}[t]
  \caption{Component and scaffold comparisons on SWE-Bench Verified with MiniMax M2.5. Single-agent variants retain graph retrieval and produce one patch without steward selection. $\Delta$ is relative to full \toolname using the same scaffold.}
  \label{tab:ablation}
  \small
  \centering
  \setlength{\tabcolsep}{4pt}
  \begin{tabular}{@{}>{\raggedright\arraybackslash}p{0.52\columnwidth}cc@{}}
  \toprule
  \textbf{Configuration} & \textbf{Rate (\%)} & \textbf{$\Delta$} \\
  \midrule
  \toolname (SWE-Agent) & \textbf{79.8} & - \\
  \toolname (mini-SWE-Agent) & 79.2 & - \\
  \midrule
  \quad w/o validation-strategy agent & 78.4 & $-1.4$ \\
  \quad w/o patch-focused change agent & 77.8 & $-2.0$ \\
  \quad w/o software-evolution graph & 79.0 & $-0.8$ \\
  \midrule
  \quad SWE-Agent & 76.4 & $-3.4$ \\
  \quad mini-SWE-Agent & 75.8 & $-3.4$ \\
  \bottomrule
  \end{tabular}
\end{table}

\subsection{Retrieval Quality and Relation Contributions}
\label{sec:graph_analysis}

\paragraph{Human evaluation of retrieved experience.}

Table~\ref{tab:recall} reports the internal expert evaluation on 100 SWE-Bench Verified tasks using the protocol in Section~\ref{sec:human_protocol}.
At the same depth of ten cards, \toolname more than doubles Useful relative to Embedding, from 1.24 to 2.87, and raises Hit from 68\% to 89\%.
Graph-grounded retrieval thus increases both the number of useful cards and the fraction of tasks receiving at least one.
Random retrieval yields 0.18 useful cards and 14\% Hit, indicating that relevant experience must be selected from the card store.

Semantic ranking finds individually similar changes; graph expansion also reaches changes connected through code, intent, or workflow evidence.
The human judgments show that this expanded search retrieves more applicable historical guidance within a fixed top-ten shortlist.
Task resolution additionally requires the agent to select and adapt that guidance to the current code. The retrieval and benchmark evaluations therefore assess successive stages of experience reuse.

Useful and Hit distinguish two aspects of retrieval quality: how many tasks receive useful experience and how much useful experience those tasks receive. At a depth of ten cards, the reported Useful values correspond to useful-card fractions of 12.4\% for Embedding and 28.7\% for \toolname. Dividing Useful by Hit gives approximately 1.82 and 3.22 useful cards, respectively, among tasks with at least one useful card. These values are derived from the same annotations. The aggregate improvement therefore combines broader coverage with more useful experience per covered task. For an agent that reads a subset of the shortlist, the higher useful-card fraction offers more task-applicable material to choose from while keeping the number of displayed candidates fixed.

\begin{table}[htbp!]
  \caption{Human evaluation of retrieval quality on 100 randomly sampled SWE-Bench Verified instances. 
Useful is the average number of useful cards among the top-10 retrieved; Hit is the fraction of instances with at least one useful experience in the top-10.}
  \label{tab:recall}
\small
\centering
\begin{tabular}{lcc}
\toprule
\textbf{Retrieval Method} & \textbf{Useful} & \textbf{Hit} \\
\midrule
Random                 & 0.18 & 14\% \\
Embedding              & 1.24 & 68\% \\
\toolname       & \textbf{2.87} & \textbf{89\%} \\
\bottomrule
\end{tabular}
\end{table}

\begin{figure}[htbp!]
  \centering
  \includegraphics[width=\linewidth]{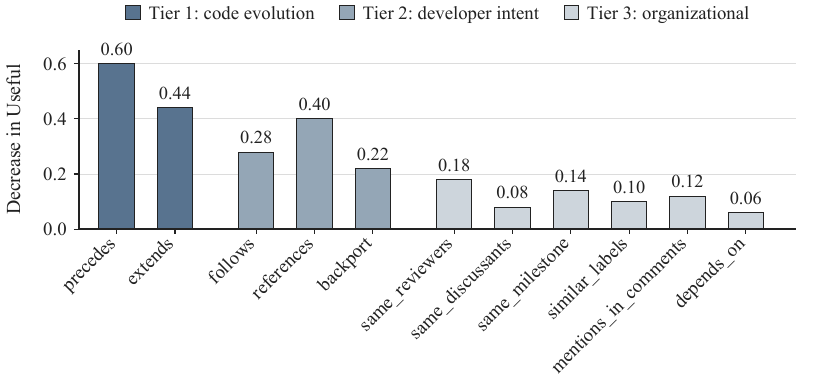}
  \caption{Per-edge-type ablation. Bars show the decrease in Useful
  (average useful cards among the top-10 retrieved) after removing each relation;
  larger drops indicate a greater loss of useful cards under the tested configuration.}
  \Description{Vertical bar chart grouped by relation tier. The decrease
  in Useful is 0.60 for precedes and 0.44 for extends in Tier 1;
  0.28 for follows, 0.40 for references, and 0.22 for backport in Tier 2;
  and 0.18 for same reviewers, 0.08 for same discussants, 0.14 for same milestone,
  0.10 for similar labels, 0.12 for mentions in comments, and 0.06 for depends on
  in Tier 3.}
  \label{fig:edge_ablation}
\end{figure}

\paragraph{Edge type contribution.}
Figure~\ref{fig:edge_ablation} removes each edge type from graph expansion and measures the decrease in Useful under the same annotation protocol.
The largest drops occur for the code-continuity relations \texttt{precedes} (0.60) and \texttt{extends} (0.44).
Developer-intent relations follow: \texttt{references} decreases Useful by 0.40, \texttt{follows} by 0.28, and \texttt{backport} by 0.22.
Organizational relations produce smaller drops, ranging from 0.06 to 0.18.
Under the tested retrieval configuration, this pattern supports prioritizing code-grounded links while using developer-intent and organizational relations to recover additional useful experience.

The removal losses also distinguish relation coverage from retrieval value. In Table~\ref{tab:edge_types}, \texttt{follows} has the broadest reported node coverage, yet removing \texttt{precedes} produces a larger loss in Useful. Coverage describes how widely a relation occurs in the stored graph, whereas removal loss measures its effect on useful cards returned for the evaluated queries. A frequent relation need not provide the most useful route for a particular task. Removing an edge type can change both which PRs are reachable and which explored paths determine their scores. Each bar therefore measures the effect of removing one relation type from the full graph; the losses are not additive shares of a fixed total. Their ordering reflects the retrieval value of the relation types in the presence of the remaining graph structure.

\section{Discussion and Threats to Validity}
\label{sec:discussion}

\subsection{Repository Experience as a Reusable Memory Layer}
The evaluated integrations show that the same repository memory can support both single-candidate agents and a workflow that generates and selects between candidates.
The same search and read interfaces serve SWE-Agent and mini-SWE-Agent, separating historical memory from the scaffold responsible for repository interaction.
Across benchmarks, the card schema, graph-construction rules, and retrieval procedure are shared, while the historical content belongs to each target repository.
The gains on issue resolution and feature implementation show the value of separating reusable development knowledge from the agents that apply it, without requiring model fine-tuning.

\subsection{What Relations Add to Experience Cards}
\label{sec:weight_robustness}
Historical relevance depends on both what a change describes and how it relates to other development work.
Connected PRs can address different symptoms or objectives and therefore occupy different positions in a semantic ranking.
Graph expansion brings their shared code, explicit references, follow-up work, and workflow context into candidate discovery.
These relations determine which additional cards the server retrieves. The agent then uses the cards' motivation and solution to judge whether the historical reasoning applies to the current code.

The increase in Useful and Hit shows the value of this organization for recovering actionable experience.
The relation-removal results favor code continuity, with developer-intent and organizational links providing additional retrieval value under the tested configuration.
The weights encode a hand-specified retrieval prior and remain unchanged across the evaluated benchmarks.
The relation-removal experiment evaluates contributions under this fixed prior; sensitivity to the numerical weight constants remains unmeasured.

\subsection{Computational Structure}
PR distillation, embedding, and graph construction incur preprocessing costs when the repository memory is built or updated.
The resulting memory can serve multiple tasks from the same repository.
Online retrieval requires query embedding and bounded graph traversal, while card inspection consumes agent context.

The retrieval parameters control different parts of this online work. At the default $K=5$, semantic search requests up to 50 index candidates before filtering, and the final shortlist contains at most ten PRs. The hop limit $H$ bounds the distance from a semantic entry, while the frontier limit $B$ bounds the number of states that continue to the next hop. Expanding a state still requires inspecting its outgoing edges, so neighborhood density affects traversal work even when $H$ and $B$ are fixed. The result limit controls how much candidate metadata reaches the agent. Subsequent read calls add full cards to the context, with the number of calls determined by the investigation. Graph exploration, shortlist size, and full-card reading therefore incur different costs, allowing the server to search more history than the agent ultimately reads.

The full resolution workflow also runs two candidate generators and invokes the steward for distinct, non-empty candidates, following Table~\ref{tab:patch_routing}.
The step and cost limits apply per agent run, so total task expenditure includes both generation runs, any steward run, and retrieval.
The single-agent variants provide an alternative execution path through the same memory.
Section~\ref{sec:ablation} reports their resolution rates under the same per-agent limits.

\subsection{Evidence Scope and Experience Quality}
\label{sec:distillation_audit}
LLM distillation of PR descriptions, discussions, and patches can omit context or introduce inaccuracies.
The human study measures task-conditioned usefulness rather than source-level factual accuracy.
Agreement between the two internal annotators is $\kappa=0.86$ before adjudication; their shared research context can introduce evaluator bias.
The prompts require agents to inspect current code and assess retrieved experience before using it.

The evaluation covers six backbones on SWE-Bench Verified and MiniMax M2.5 on the two transfer benchmarks.
Repeated-run statistics describe the full MiniMax configuration; the baseline and remaining configurations use single runs.
The single-agent variants combine graph retrieval with role-specific prompting, so their gains over SWE-Agent reflect that combination.
Removing a strategy also removes the second candidate and steward selection, so the full workflow's additional gain reflects their combined effect.
Historical retrieval is restricted by recorded PR creation and merge timestamps as described in Section~\ref{sec:setup}.

\subsection{Implications for Software Maintenance}
\label{sec:broader_impacts}
Repository history records implementation patterns, constraints, and rationales that are often revisited during maintenance.
\toolname makes these records searchable when an agent develops or reviews a new change.
Retrieved experience supplies hypotheses and constraints for investigating current code, connecting repository-history mining to automated code modification.
Memory contributes prior development knowledge to this process, while tests and code review assess the resulting patches.

The reusable unit is a development decision together with the conditions that motivated it. A historical Solution describes an implementation in an earlier software state, while Motivation explains the behavioral gap or assumption that made the implementation appropriate. Reading them together helps an agent determine which parts of the experience transfer to the current code. Relations connect that decision to subsequent development on the same files or to explicitly linked work, extending the historical context available for inspection. Repository history thus becomes a connected record of evolving requirements and implementation choices. This representation supports reasoning about the constraints a change must preserve and the rationale for modifying the code, as well as the concrete implementation itself.

\section{Related Work}

\subsection{Software Engineering Agents and Benchmarks}

Automated program repair has progressed from search-based methods~\cite{le2012genprog,kim2013automatic} and sequence-to-sequence models~\cite{chen2019sequencer,lutellier2020coconut} to interactive LLM-based repair. Feedback-driven methods~\cite{yin2024thinkrepair,xia2024automated} use execution signals across turns, while ReduceFix~\cite{yang2026reducefix} reduces test inputs while preserving failure-inducing behavior to guide patch generation. AutoCodeRover~\cite{zhang2024autocoderover}, OpenHands~\cite{wangopenhands}, and Trae Agent~\cite{gao2025trae} support repository exploration and code modification. PlayCoder~\cite{peng2026playcoder} combines repository-aware GUI code generation with interactive testing and iterative repair. Agentless~\cite{xia2024agentless} simplifies the workflow, while SWE-RL~\cite{wei2025swerl} and Kimi~\cite{team2025kimi} improve models through reinforcement-learning-based post-training.

SWE-Bench~\cite{jimenez2024swebench} evaluates repository-level issue resolution, while SWE-Bench Pro~\cite{deng2025swe}, SWE-PolyBench~\cite{rashid2025swe}, Multi-SWE-bench~\cite{zan2025multi}, and SWE-rebench~\cite{badertdinov2025swe} extend evaluation to harder tasks, multiple programming languages, or continually collected instances. 
Other benchmarks cover feature implementation~\cite{li2025fea,zhoufeaturebench}, migration~\cite{liu2025migrationbench,cheng2025codemenv}, and refactoring~\cite{gautamrefactorbench,thillen2026codetaste}. 
\toolname complements these agents with historical PR experience for candidate generation and selection.

\subsection{Knowledge Reuse for Software Engineering Agents}

Retrieval-augmented generation~\cite{lewis2020retrieval} supplies code tasks with retrieved snippets, API usages, documentation, or repository fragments~\cite{parvez2021retrieval,lu2022reacc,zhou2022docprompting,zhang2023repocoder,ding2023crosscodeeval,wang2024coderag}.
Graph-based retrieval organizes source material through relations~\cite{edge2024graphrag,hu2024grag,mostafa2024g}, while agent-memory systems reuse episodic records, skills, and reflections across interactions~\cite{park2023generative,zhong2024memorybank,packer2023memgpt,wang2023voyager,shinn2023reflexion,zhao2024expel}.

Code graphs represent program structure for repository exploration and generation.
LocAgent~\cite{chen2025locagent} navigates files, classes, and functions through dependency relations for code localization.
GraphCoder~\cite{liu2024graphcoder} combines control flow and control/data dependence in statement-level graphs to retrieve code contexts for completion.
CGM~\cite{tao2025cgm} encodes code-entity semantics as node tokens and uses graph adjacency to constrain attention in a trained model.
RepoDistill~\cite{yin2026repodistill} combines graph-based retrieval with learned token-budget allocation to compress repository context.
\toolname uses a graph of historical PR events to discover experience about prior changes, with the retrieved cards exposed through task-time tools.

Systems that reuse software-engineering history draw on several sources of experience.
SWE-Exp~\cite{chen2025sweexp} and ExpeRepair~\cite{mu2025experepair} reuse experience from agents' own prior trajectories.
MemGovern~\cite{wang2026memgovern} pools human repair experience from issues and merged patches across GitHub repositories.
HAFix~\cite{shi2025hafix} uses blame-derived context, Repository Memory~\cite{wang2025repomem} retrieves commits and linked issues, Lingxi~\cite{yang2025lingxi} extracts procedural knowledge from earlier issues, and MemCoder~\cite{deng2026memcoder} builds intent-to-code memory from commits in the target repository.
\toolname likewise draws on development history from the target repository and organizes retrieval around relations between historical PR events.

MemGovern~\cite{wang2026memgovern} is particularly close in its use of structured cards and a search-then-browse interface.
Its index layer captures problem summaries and diagnostic signals, while its resolution layer stores repair reasoning.
Search ranks cards by query-to-index embedding similarity, and the agent can revise queries and selectively browse their detailed contents.
\toolname retains selective card reading and adds weighted expansion from semantic entry points over typed relations between PRs.
Changed files, explicit references, follow-up work, reviewers, milestones, and other PR metadata provide the evidence for these links.
This changes candidate discovery: a card can be retrieved through its connection to a relevant historical change as well as its own query similarity.

Repository Memory~\cite{wang2025repomem} augments LocAgent with BM25 search over commit messages, inspection of patches and linked issues, and functionality summaries of frequently edited files.
Its memory tools complement LocAgent's navigation of the current code graph.
\toolname builds its retrieval graph over historical PR events, using links between changes to retrieve the experience associated with those events.
The resulting context is available during both candidate generation and history-informed patch selection.

\section{Conclusion}

We presented \toolname, a test-time framework that turns PR history into graph-structured software-evolution experience for code agents.
Its central contribution is relational retrieval: semantic search locates relevant historical changes, and typed PR relations make connected experience available for selective reading.
Structured cards supply the rationale and implementation logic used during candidate generation and patch selection.

Across six backbones on SWE-Bench Verified, the full workflow raises the mean SWE-Agent resolution rate from 69.2\% to 72.9\% and reaches 79.8\% in the strongest configuration.
With MiniMax M2.5, it raises resolution rates from 48.3\% to 51.7\% on SWE-Bench Pro and from 41.0\% to 43.5\% on FEA-Bench Lite.
The experience-guided single-agent variants also improve on the base agent.
With ten cards retrieved per task, graph-grounded retrieval raises the average number of human-judged useful cards from 1.24 to 2.87 over flat semantic retrieval.
These findings establish PR relations as a useful retrieval signal for repository memory and show how that memory can support both patch generation and selection in the evaluated workflows.

\clearpage

\bibliographystyle{ACM-Reference-Format}
\bibliography{reference}

\end{document}